\documentclass[showpacs,preprintnumbers,twocolumn,amsmath,amssymb,floatfix,,prd]{revtex4}

\usepackage{graphicx}
\usepackage{dcolumn}
\usepackage{bm}
\usepackage{hyperref}
\usepackage{color}

\begin{document}

\title{A generalized Finch-Skea class one static solution}

\author{Ksh. Newton Singh}
\email{ntnphy@gmail.com} 
\affiliation{Department  of  Physics,  National  Defence  Academy, Khadakwasla,  Pune-411023,  India\\
Department  of  Mathematics, Jadavpur  University,  Kolkata 700032,  India}

\author{S. K. Maurya}
\email{sunil@unizwa.edu.om,} 
\affiliation{Department of Mathematical and Physical Sciences, College of Arts and Science, University of Nizwa, Nizwa, Sultanate of Oman}

\author{Farook Rahaman}
\email{rahaman@associates.iucaa.in,} 
\affiliation{Department of Mathematics, Jadavpur University, Kolkata 700032, West Bengal, India}

\author{Francisco Tello-Ortiz} \email{francisco.tello@ua.cl}
\affiliation{Departamento de F\'isica, 
Facultad de ciencias b\'asicas, Universidad de Antofagasta, Casilla 170, 
Antofagasta, Chile}

\date{\today}

\begin{abstract}
In the present article, we discuss relativistic anisotropic solutions of the Einstein field equation for the spherically symmetric line element under the class one condition. To do so we apply the embedding class one technique using Eisland condition. Within this approach, one arrives at a particular differential equation that links the two metric components $e^{\nu}$ and $e^{\lambda}$. In order to obtain the full space-time description inside the stellar configuration we ansatz the generalized form of metric component $g_{rr}$ corresponding to the Finch-Skea solution. Once the space-time geometry is specified we obtain the complete thermodynamic description i.e. the matter density $\rho$, the radial, and tangential pressures $p_r$ and $p_t$, respectively.  Graphical analysis shows that the obtained model respects the physical and mathematical requirements that all ultra-high dense collapsed structures must obey. These salient features concern well behaved metric potentials and thermodynamic observables i.e. free from geometrical and physical singularities within the object, preservation of causality in both radial and tangential direction, well behaved and positively defined energy-momentum tensor and , stability trough Abreu, adiabatic index and Harrison-Zeldovich-Novikov criteria, anisotropy factor, compactness factor and equilibrium via modified Tolman-Oppenheimer-Volkoff equation. The $M-R$ diagram suggest that the solution yields stiffer EoS as parameter $n$ increases. The $M-I$ graph is in agreement with the concepts of Bejgar et al. \cite{bej} that the mass at $I_{max}$ is lesser by few percent (for this solution $\sim 3\%$) from $M_{max}$. This suggest that the EoSs is without any strong high-density softening due to hyperonization or phase transition to an exotic state.

\end{abstract}


\maketitle

\section{Introduction}\label{Sec1}

It is well known that Einstein’s general theory of relativity has fruitfully explained about several observations or cosmological measures including astrophysical backgrounds\cite{Tipler,Shapiro}. The golden age of cosmology saw the theory of Hubble, the material, the biological structure, the nuclear synthesis, as well as the higher level of precision in explaining the potential origin of the universe and its subsequent evolution.  Basically Einstein general theory of relativity is generalization of Newtonian gravity which is mainly suitable to describe the structure of compact star in the strong gravitational fields. Few of these compact objects like pulsars, black holes and neutron stars have densities of the order greater than or equal $10^{14}gm/cm^3$. The Schwarzschild has discovered the first precise solution of Einstein field equations for the gravitational field in the inner part of a non-circular spherical body consisting of a non-compressible fluid. This is also known as constant density solution with outer being empty and has zero pressure at the surface. Now a days, the researcher are busy on the study of relativistic compact stars.  For object modeling, we study the solutions of Einstein's equations of static spherically symmetric with different physical causes. These solution may be stated as perfect fluid, anisotropic fluid, and dust.  However, there are strong theoretical evidence that steep excessive dense celestial bodies are not made of perfect fluids. In some cases, the objects with different physical phenomena are found, for example anisotropy.  The first theoretical attempt to look at the effect of variance was seen in about 1922 when Jeans \cite{Jeans} looked anisotropic pressure on the self-gravitating bodies of Newtonian configurations.\\

After this, Ruderman \cite{Ruderman} has studied the effect of the anisotropy. He said that the stars may have anisotropic characteristics at very high density of the order $10^{15} gm/cm^3$ where the nuclear interaction becomes relativistic. Sudden after, Bowers and Liang\cite{Bowers} studied about confined properties of relativistic anisotropic matter distribution for static spherically symmetric configurations, which is comprehensively populated. Recently, An extensive research was conducted in the study of physics related to anisotropic pressures.In this connection, Dev and Gleiser \cite{Dev1,Dev2} have shown that pressure variation affects the physical properties of mass, structure and excessive pressure areas. Also there are other several analytical static solutions have been already discovered by several authors \cite{Herrera1985,Maurya1,Maurya2,Maurya3,Singh1,Singh2,Maurya4,Maurya5,Deb1,Maurya6,Maurya7,Maurya8,Singh3,Deb2,Maurya9,Maurya10,Maurya11,Maurya12,Maurya13,Maurya14,Gupta1,Mak2003}. Most pioneering work by Herrera and Santos \cite{Herrera1} where they have specified about effect of local anisotropy in self gravitating systems.  More remarkably, the algorithm for all possible static isotropic, anisotropic as well as charged anisotropic solutions of Einstein's equation for the spherically symmetric line element can be attractively determined by a general procedure which are given in Refs. \cite{Lake,Herrera2,Maurya2017}. \\

It is essential to note that the redshift and mass of the stellar model both varies with the anisotropy. Recently, an extensive efforts have been made in the modeling of physical observed astronomical objects in the existence of anisotropy which can be seen in recent research papers\cite{Sharma,Ngubelanga,Murad1,Murad2} and the references therein. In these recent papers, the physical analysis reaffirms the significance of the presence of a non-zero anisotropy in the modeling of astrophysical objects. In order to create a substantially reliable object, it is necessary to find an analytical solution of Einstein field equations for relativistic matter distribution which can be solved by restricting the space-time geometry or stating an equation of state (EOS) of the matter distribution.
On the other hand, we can generate the exact solution of relativistic field equation using another different approach known as embedding class one condition.  In this connection, Riemann has presented the idea, known as Riemannian geometry, to study the essential geometric properties of the objects.  Immediately after this, Schlaefli \cite{Schlaefli} estimated that a Riemannian manifold of metric which is analytic with positively defined signature can be embedded locally and isometrically into the higher dimensional flat Euclidean space. 

The idea of embedding locally and isometrically an n-dimensional Riemannian manifold $V_n$ into an $N = n(n + 1)/2$ dimensional pseudo-Euclidean space was proved in the past by authors \cite{Janet,Cartan,Burstin}. The embedding class $p$ of $V_n$ is the minimum number of extra dimensions required by the pseudo-Euclidean space, which is obviously equal to $p=N-n = n (n-1) /2$.  As we know, general theory of relativity deals only with four dimensional spacetime, however embedding class solution may provide new characteristics to gravitational field, as well to physics. In case of relativistic space time $V_n$, the embedding class $p$ turns out to be $p=6$. In particular the classes of spherical and plane symmetric space-time are $p=2$ and $p=3$ respectively. The famous Friedman-Robertson-Lemaitre space-time, is of class $p=1$, while the Schwarzschild’s exterior and interior solutions are of class $p=2$ and class $p=1$ respectively, moreover Kerr metric is class 5.  In the literature \cite{Barnes1,Kumar,Barnes2,Ponce,Akbar,Abbas,Kuhfitting1,Kuhfitting2}, there are many interesting work concerning  the effects of the technique of embedding of lower dimensional Riemannian space into the higher dimensional pseudo-Euclidean space in the framework of GR.  The main consequence of embedding a Riemannian variety corresponding to a spherically symmetric and static spacetime into a pseudo Euclidean space is the so-called Eisland condition. This condition links both metric potentials $e^{\nu}$ and $e^{\lambda}$ into a single differential equation. It is a mathematical simplification which reduces the problem of obtaining exact solutions to a single-generating function. The approach is to choose one of the gravitational potentials on physical grounds and to then integrate the Eisland condition to fully specify the gravitational behavior of the model. In this paper we utilize Eisland condition to derive solutions which describe compact objects in general relativity. We subject our solutions to rigorous physical tests which ensure that they do describe physically observable objects in the universe. 
The article is organized as follows:  In Sec. II we have specified the interior space time and Einstein field equations for anisotropic matter distribution. This section also includes the embedding class one condition along with non-vanishing Riemannian tensor for interior space time. IN next section III, we have presented a generalized Finch-Skea solution for anisotropic matter distribution using the class one condition. The nonsingular nature of pressures, density and bounds of the constant are given in Sec. IV. In Sec. V, we presents the necessary and sufficient conditions to determine all possible constant parameters that describe the anisotropic solution. For this purpose, we match our interior space-time to the exterior space-time (Schwarzscild metric). The section VI includes the energy conditions. In Sec. VII, we have discussed the most important features of the objects like equilibrium condition via. Tolman-Oppenhimer-Volkoff equation, Causality and stability condition through Herrera Aberu criterion, adiabatic index and Harrison-Zeldovich-Novikov static stability criterion…… 

\section{Interior space-time and field equations}

The interior space-time for spherically symmetric space-time is chosen as,
\begin{equation}
ds^{2}=e^{\nu(r)}dt^{2}-e^{\lambda(r)}dr^{2}-r^{2}\left(d\theta^{2}+\sin^{2}\theta d\phi^{2} \right) \label{met}
\end{equation}
where $\nu$ and $\lambda$ are functions of the radial coordinate `$r$' only.\\

 The Einstein's field equations corresponding an anisotropic fluid distribution becomes
\begin{eqnarray}
R^\mu_\nu-{1\over 2}g^\mu_\nu R &=& -{8\pi} \big[(p_t +\rho c^2)v^\mu v_\nu-p_t g^\mu_\nu+(p_r-p_t) \nonumber \\
&& \chi_\nu \chi^\mu \big] \label{fil}
\end{eqnarray}
where the symbols have their usual meanings.\\

For the space-time \eqref{met}, the field equations can be written as
\begin{eqnarray}
\frac{1-e^{-\lambda}}{r^{2}}+\frac{e^{-\lambda}\lambda'}{r} &=& 8\pi\rho \label{dens}\\
\frac{e^{-\lambda}-1}{r^{2}}+\frac{e^{-\lambda}\nu'}{r} &=& 8\pi p_{r} \label{prs}\\
e^{-\lambda}\left(\frac{\nu''}{2}+\frac{\nu'^{2}}{4}-\frac{\nu'\lambda'}{4}+\frac{\nu'-\lambda'}{2r} \right) &=& 8\pi p_t. \label{prt}
\end{eqnarray}

The measure of anisotropy is defined as $\Delta = 8\pi (p_t - p_r)$.\\
On the other hand, It was proved by Eisenhart~\cite{Eisenhart1925} that an embedding class 1 space (A $(n+1)$ dimensional space $V^{n+1}$ can be embedded into a $(n+2)$ dimensional pseudo-Euclidean space $E^{n+2}$) can be described by a $(n+1)$ dimensional space $V^{n+1}$ if there exists a symmetric tensor $a_{mn}$ which satisfies the following Gauss- Codazzi equations: \\
\begin{eqnarray}\label{eqcls1.1}
R_{mnpq}=2\,e\,{a_{m\,[p}}{a_{q]n}}~~~\nonumber\\ \text{and}~~~a_{m\left[n;p\right]}-{\Gamma}^q_{\left[n\,p\right]}\,a_{mq}+{{\Gamma}^q_{m}}\,{}_{[n}\,a_{p]q}=0,
\end{eqnarray}

where $e=\pm1$, $R_{mnpq}$ denotes the curvature tensor and square brackets represent antisymmetrization. Here, $a_{mn}$ are the coefficients of the second differential form. Moerover, A necessary and sufficient condition for the embedding class I of Eq.~\ref{eqcls1.1} in a suitable convenient form was given by Eiesland \cite{Eiesland1925} as


 \begin{eqnarray}
R_{{0101}}R_{{2323}}=R_{{0202}}R_{{1313}}-R_{{1202}}R_{{1303}}.\label{3.2}
\end{eqnarray}

The non-vanishing components of Riemannian tensor for the  spherically symmetric interior space-time (\ref{met}) are given as 

\begin{eqnarray}
&& R_{{0101}}=-\frac{1}{4}\,{{\rm e}^{\nu}} \left( -\nu^{{\prime}}\lambda^{{\prime}}+{\nu^{{\prime}}}^{2}+2
\,\nu^{{\prime\prime}} \right), \nonumber \\
&& R_{{2323}}=-{r}^{2} {\sin^2 \theta} \left( 1-{{\rm e}^{-\lambda}} \right),~~
  R_{{0202}}=-\frac{1}{2}\,r\nu^{{\prime}}{{\rm e}^{\nu-\lambda}},\nonumber\\
&& R_{{1313}}=-\frac{1}{2}\,\lambda^{{\prime}}r \sin^2 \theta,~~ R_{{1202}}=0,~~~ R_{{1303}}=0
\end{eqnarray}
  \\
  
By plugging the values of above Riemannian components into Eq.~(\ref{3.2}) we obtain a differential equation in $\nu$ and $\lambda$ of the form 
\begin{eqnarray}\label{3.3}
({\lambda}^{{\prime}}-{{\nu}^{{\prime}
}})\,{\nu}^{{\prime}}\,{{\rm e}^{\lambda}}+2\,(1-{{\rm e}^{\lambda}}){\nu}^{{\prime\prime}}+{{\nu}^{{\prime}}}^{2}=0.
\end{eqnarray}

The solutions  Eq.(\ref{3.3}) of are named as `embedding class one solution" and they can be embedded in five dimensional pseudo-Euclidean space.\\

On integration of Eq.(\ref{3.3}) we get
\begin{equation}
e^{\nu}=\left(A+B\int \sqrt{e^{\lambda}-1}~dr\right)^2\label{nu1}
\end{equation}
where $A$ and $B$ are constants of integration.

By using (\ref{nu1}) we can express the anisotropy as \cite{ma1,ma2} 
\begin{eqnarray}
\Delta = {\nu' \over 4e^\lambda}\left[{2\over r}-{\lambda' \over e^\lambda-1}\right]~\left[{\nu' e^\nu \over 2rB^2}-1\right]. \label{del1}
\end{eqnarray}

For isotropic case $\Delta=0$ and there are three possible solutions when (a) $e^\nu = C$ and $e^\lambda=1$ (not physical), (b) Schwarzschild interior solution (not physical) and (c) Kohler-Chao solution (cosmological solution as the pressure vanishes at $r\rightarrow \infty$).

\section{A generalized solution for compact star model}

\begin{figure}[t]
\centering
\resizebox{0.8\hsize}{!}{\includegraphics[width=7cm,height=4.5cm]{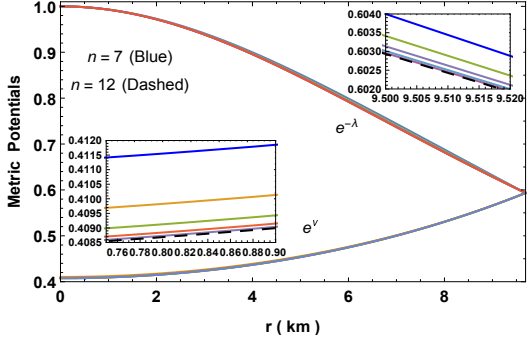}}
\caption{Variation of metric potentials w.r.t radial coordinate $r$ for $M = 1.97 M_{\odot},R = 9.69 km$ and $b = 0.04$.}\label{mt}
\end{figure}

Since the field equations depend on metric functions $\nu$ and $\lambda$. To construct a viable anisotropic model, We have assumed the generalized form of Finch-Skea metric \cite{Finch} function $g_{rr}$ as
\begin{eqnarray}
\lambda &=& \ln (1+a r^2+b^{n-1} r^n)\label{ela}
\end{eqnarray}
 where $a$ and $b$ are non-zero positive constants and $n$ is a positive integer. By substituting the value of $\lambda$ from Eq.\ref{ela} into Eq.(\ref{nu1}) we get
\begin{eqnarray}
e^\nu &=& \bigg(A-\bigg\{2 B \Big[a b (n-2) r^2 f(r) \sqrt{a b^{1-n} r^{2-n}+1}+\nonumber \\
&& (6-n) \left(a b r^2+b^n r^n\right)\Big]\bigg\} {(a+b^{n-1} r^{n-2})^{-1/2} \over b (n-6) (n+2)}\bigg)^2 \nonumber \\ \label{enu}
\end{eqnarray}
where $f(r) = ~_2F_1\left(\frac{1}{2},\frac{n-6}{2 (n-2)};\frac{10-3 n}{4-2 n};-a b^{1-n} r^{2-n}\right)$ is known as Gauss hypergeometric function. The behaviour of the metric potentials are plotted in Fig. \ref{mt}.\\

\begin{figure}[t]
\centering
\resizebox{0.8\hsize}{!}{\includegraphics[width=7cm,height=4.5cm]{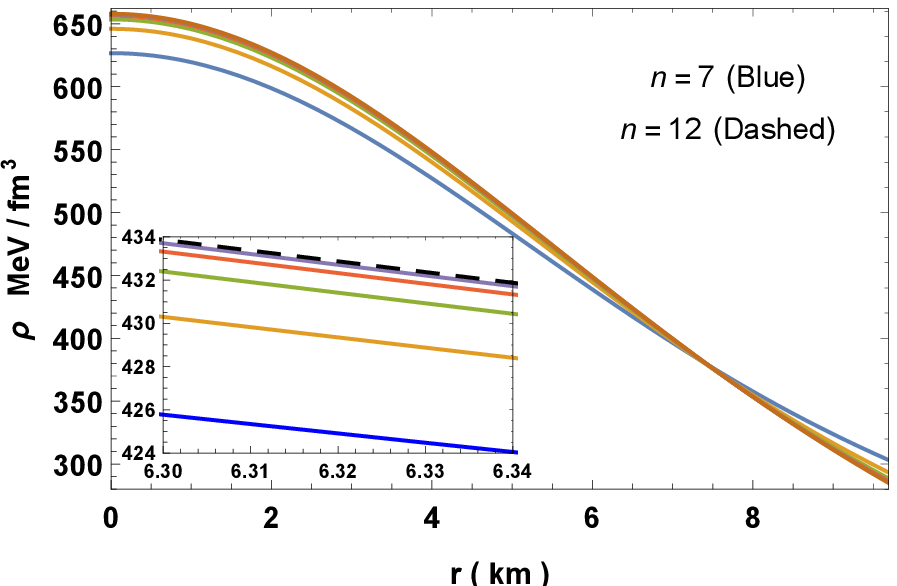}}
\caption{Density profile of PSR J1614-2230 for $M = 1.97 M_{\odot},R = 9.69 km$ and $b = 0.04$.}\label{fid}
\end{figure}

\begin{figure}[t]
\centering
\resizebox{0.8\hsize}{!}{\includegraphics[width=7cm,height=4.5cm]{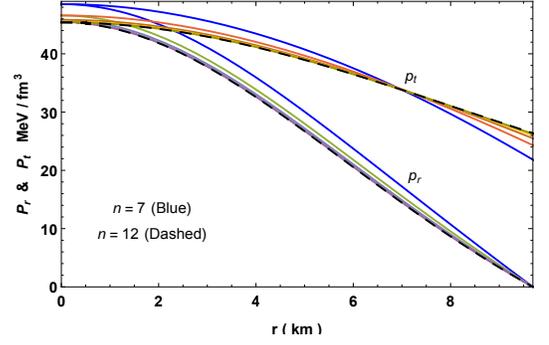}}
\caption{Radial and transverse pressure profile of PSR J1614-2230 for $M = 1.97 M_{\odot},R = 9.69 km$ and $b = 0.04$.}\label{fip}
\end{figure}

\begin{figure}[t]
\centering
\resizebox{0.8\hsize}{!}{\includegraphics[width=7cm,height=4.5cm]{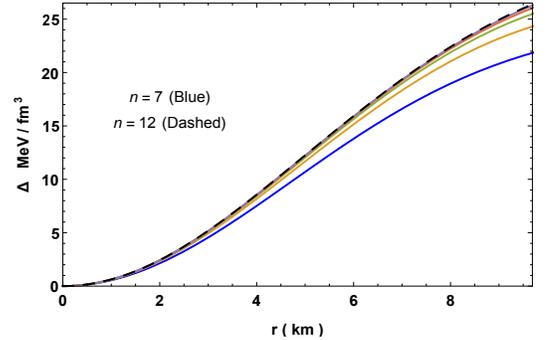}}
\caption{Anisotropy profile of PSR J1614-2230 for $M = 1.97 M_{\odot},R = 9.69 km$ and $b = 0.04$.}\label{fia}
\end{figure}

\begin{figure}[t]
\centering
\resizebox{0.8\hsize}{!}{\includegraphics[width=7cm,height=4.5cm]{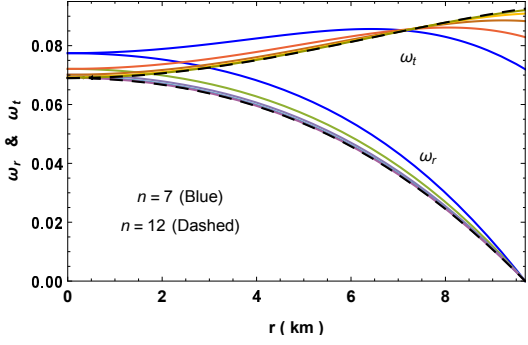}}
\caption{Equation of state parameter profiles of PSR J1614-2230 for $M = 1.97 M_{\odot},R = 9.69 km$ and $b = 0.04$.}\label{fie}
\end{figure}

By using the metric potentials $\nu$ and $\lambda$, we directly obtain the expression for thermodynamic variables like density, radial and transverse pressure and anisotropy as
\begin{eqnarray}
8\pi \rho(r) &=& \frac{1}{\left(a r^2+b^{n-1} r^n+1\right)^2} \bigg[a^2 r^2+a \left(2 b^{n-1} r^n+3\right)\nonumber \\
&& +b^{n-1} r^{n-2} \left(b^{n-1} r^n+n+1\right)\bigg]\label{den}\\
8\pi  p_r(r) &=& \Big[(n-6) b^n k(r) r^n \Big\{b \Big[2 B r \left(a r^2-n-2\right)+\nonumber \\
&& A (n+2) j(r)\Big]+2 B b^n r^{n+1}\Big\}-2 a b B (n-2) \nonumber \\
&& r^3 f(r) (a b r^2+b^n r^n) \Big] \Big[(6-n) \Big\{2 a b B r^3+A b \nonumber \\
&& (n+2) j(r)+2 B b^n r^{n+1}\Big\}+2 a b B (n-2) r^3 \nonumber \\
&& f(r) k(r)\Big]^{-1} \times \frac{b^{-n} r^{-n-2} \left(a b r^2+b^n r^n\right)}{k(r) \left(a b r^2+b^n r^n+b\right)}\label{pre1}\\
\Delta(r) &=& \frac{k(r) l(r) q(r)}{2 r^2 p(r) \left(a b r^2+b^n r^n\right) \left(a b r^2+b^n r^n+b\right)^2}\\
8\pi p_t (r) &=& 8\pi p_r+\Delta.
\end{eqnarray}
where,
\begin{eqnarray}
j(r) &=& \sqrt{a r^2+b^{n-1} r^n} \\
k(r) &=& \sqrt{a b^{1-n} r^{2-n}+1} \\
l(r) &=& 2 a^2 b^2 r^4+4 a b^{n+1} r^{n+2}+b^n r^n \big[2 b^n r^n\nonumber \\
&& +b (2-n)\big]\\
n(r) &=& b \left[B r (2 a r^2-n-2)+A (n+2) j(r)\right] \nonumber \\
&& +2 B b^n r^{n+1}\\
q(r) &=& 2 a b B (2-n) r^3 f(r) \left[a b r^2+b^n r^n\right]+(n-6) b^n \nonumber \\
&& k(r) n(r) r^n\\
p(r) &=& (n-6) \left[2 a b B r^3+A b (n+2) j(r)+2 B b^n r^{n+1}\right] \nonumber \\
&& +2 a b B (2-n) r^3 f(r) k(r)
\end{eqnarray}
There variations of the above physical quantities are given in Figs. \ref{fid}-\ref{fia}. We should ensure that values of $p_r/\rho$ and $p_t/\rho$ at the interior must be less than unity for a physical system (Fig. \ref{fie}).

The other physical parameters mass, compactness factor and red-shift can be determine as
\begin{eqnarray}
m(r) &=& 4\pi \int r^2 \rho ~dr=\frac{r}{2} \left(1-\frac{b}{a b r^2+b^n r^n+b}\right)~~\\
u(r) &=& {2m(r) \over r}= 1-\frac{b}{a b r^2+b^n r^n+b}\\
z(r) &=& e^{-\nu/2}-1.
\end{eqnarray}
We have plotted the $M-R$ diagram in Fig. \ref{fim}. Here we have determined the radius from surface density and determine the mass using this radius using the boundary condition. The trend of red-shift is plotted in Fig. \ref{fir}.

\section{Non-singular nature of the solution}

To check the  physical validity of the solution, we ensure that the central values of pressure and density must be finite i.e.
\begin{eqnarray}
\rho_c &=& {3 a \over 8\pi}  >0, \label{rhc}\\
p_{rc} &=& p_{tc} =  \frac{\sqrt{a} \left(2 B-\sqrt{a} A\right)}{8\pi A} > 0. \label{pc}
\end{eqnarray}

It is also require to ensure that any physical fluid satisfies the Zeldovich's criterion i.e. $p_{rc}/ \rho_c \le 1$ which implies
\begin{eqnarray}
{p_{rc} \over \rho_c} = \frac{2 B-\sqrt{a} A}{3\sqrt{a}A} \le 1. \label{zel}
\end{eqnarray}

Now a physical constraint on $B/A$ arises due to (\ref{pc}) and (\ref{zel}) as
\begin{eqnarray}
{\sqrt{2} \over a} < {B \over A} \le {2\sqrt{a}}. \label{zell}
\end{eqnarray}

\section{Boundary Conditions and determination of constants}

It is necessary that we should match our interior space-time to the exterior $Schwarzschild$ \cite{kar} line element 
\begin{eqnarray}
ds^{2} &=& \left(1-\frac{2m}{r}\right)dt^{2}-\left(1-\frac{2m}{r}\right)^{-1}dr^{2} \nonumber \\
&& -r^{2}\big(d\theta^{2}+\sin^{2}\theta d\phi^{2} \big)
\end{eqnarray}
at the boundary $r=R$. Also, the radial coordinate $r$ must be greater than $2m$ so that it doesn't form a black hole.\\ 

Using the continuity of the metric coefficients $e^{\nu}$ and $e^{\lambda}$ across the boundary ($r=R$) and vanishing of radial pressure at the boundary ($r=R$) we get the following equations  
\begin{eqnarray}
1-\frac{2M}{R} &=& e^{\nu_s} = e^{-\lambda_s}\label{b1}\\
p_r(r=R) &=& 0. \label{b3}
\end{eqnarray}

On using the boundary conditions (\ref{b1}) and (\ref{b3}) we obtain the value of arbitrary constants as, 
\begin{eqnarray}
a &=& \frac{b^n (R-2 M) R^n-2 b M}{b R^2 (2 M-R)} \\
A &=& \sqrt{1-\frac{2 M}{R}}+\frac{2 BR^2}{b (n-6) (n+2)}\Big[b (6-n) \nonumber \\
&&  \hspace{-1 mm}\sqrt{a+b^{n-1} R^{n-2}}+a (n-2) b^{\frac{3-n}{2}} f(R) R^{\frac{2-n}{2}} \Big]\\
B &=& \sqrt{1-\frac{2 M}{R}} ~\frac{b (6-n) (n+2) \sqrt{a +b^{n-1} R^{n-2}}}{2} \nonumber \\
&& \bigg[2 (n-6) b^n R^n+b (n-6) \left(a R^2-n-2\right)- \nonumber
\end{eqnarray}
\begin{eqnarray}
&& \frac{a (n-2) b^{1-n} f(R) R^{2-n} \left(a b R^2+b^n R^n\right)}{\sqrt{a b^{1-n} R^{2-n}+1}}+\nonumber \\
&& a b (n-2) R^2 f(R) \sqrt{a b^{1-n} R^{2-n}+1}+(6-n) \nonumber \\
&& \left(a b R^2+b^n R^n\right)\bigg]^{-1}
\end{eqnarray}
Here $M$ and $R$ are chosen from observed values of compact stars and $b$ as free parameter.

\section{Energy Conditions}

In this section we are willing to verify the energy conditions namely null energy condition (NEC), dominant energy condition (DEC) and weak energy condition(WEC) at all points in the interior of a star which will be satisfied if the following inequalities hold simultaneously:
\begin{eqnarray}
\text{WEC} &:& T_{\mu \nu}t^\mu t^\nu \ge 0~\mbox{or}~\rho \geq  0,~\rho+p_i \ge 0 \\
\text{NEC} &:& T_{\mu \nu}l^\mu l^\nu \ge 0~\mbox{or}~ \rho+p_i \geq  0\\
\text{DEC} &:& T_{\mu \nu}t^\mu t^\nu \ge 0 ~\mbox{or}~ \rho \ge |p_i|\\
&& \mbox{where}~~T^{\mu \nu}t_\mu \in \mbox{nonspace-like vector} \nonumber \\
\text{SEC} &:& T_{\mu \nu}t^\mu t^\nu - {1 \over 2} T^\lambda_\lambda t^\sigma t_\sigma \ge 0 ~\mbox{or}~ \rho+\sum_i p_i \ge 0. \nonumber \\
\end{eqnarray}
where $i\equiv (radial~r, transverse ~t),~t^\mu$ and $l^\mu$ are time-like vector and null vector respectively. \\ 

We will check the energy conditions with the help of graphical representation. In Fig. \ref{fiec}, we have plotted the L.H.S of the above inequalities which verifies that  all the energy conditions are satisfied at the stellar interior.

\section{Stability and equilibrium of the model }

\subsection{Equilibrium under various forces}
Equilibrium state under three forces $viz$ gravitational, hydrostatics and anisotropic forces can be analyze whether they satisfy the generalized Tolman-Oppenheimer-Volkoff (TOV) equation or not and it is given by
\begin{equation}
-\frac{M_g(r)(\rho+p_r)}{r}e^{\frac{\nu-\lambda}{2}}-\frac{dp_r}{dr}+\frac{2}{r}(p_t-p_r)=0, \label{to1}
\end{equation}
where $M_g(r) $ represents the gravitational mass within the radius $r$, which can derived from the Tolman-Whittaker formula and the Einstein field equations and is defined by

\begin{eqnarray}
M_g(r) &=& 4 \pi \int_0^r \big(T^t_t-T^r_r-T^\theta_\theta-T^\phi_\phi \big) r^2 e^{\nu+\lambda \over 2}dr .\label{mg}
\end{eqnarray}

For the Eqs. (\ref{dens})-(\ref{prt}), the above Eq. (\ref{mg}) reduced to
\begin{equation}
M_g(r)=\frac{1}{2}re^{(\lambda-\nu)/2}~\nu'.
\end{equation}

Plugging the value of $M_g(r)$ in equation (\ref{to1}), we get
\begin{equation}
-\frac{\nu'}{2}(\rho+p_r)-\frac{dp_r}{dr}+\frac{2}{r}(p_t-p_r)=0.
\end{equation}

\begin{figure}[t]
\centering
\resizebox{0.8\hsize}{!}{\includegraphics[width=7cm,height=4.5cm]{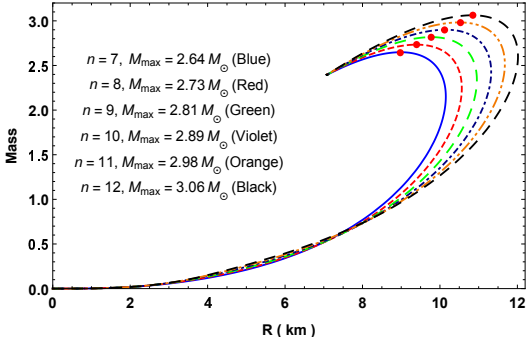}}
\caption{M-R diagram for $a=0.001$ and $b = 0.04$.}\label{fim}
\end{figure}

\begin{figure}[t]
\centering
\resizebox{0.8\hsize}{!}{\includegraphics[width=7cm,height=4.5cm]{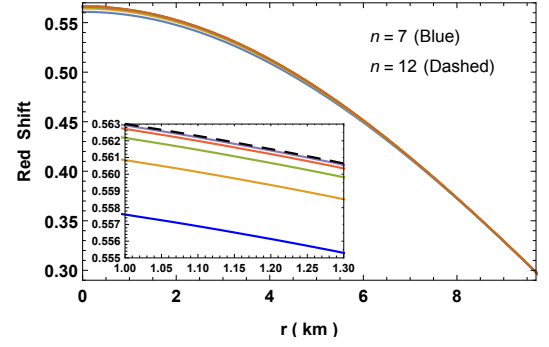}}
\caption{Red-shift profiles of PSR J1614-2230 for $M = 1.97 M_{\odot},R = 9.69 km$ and $b = 0.04$.}\label{fir}
\end{figure}

\begin{figure}[t]
\centering
\resizebox{0.8\hsize}{!}{\includegraphics[width=7cm,height=4.5cm]{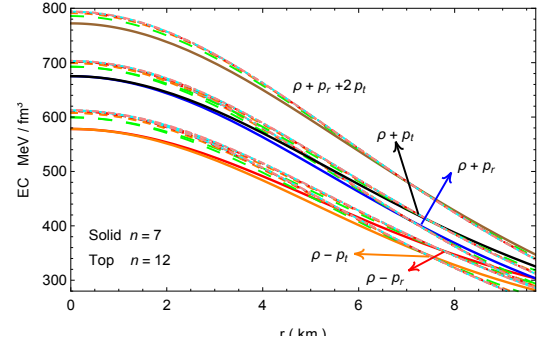}}
\caption{Energy Consitions of PSR J1614-2230 for $M = 1.97 M_{\odot},R = 9.69 km$ and $b = 0.04$.}\label{fiec}
\end{figure}

\begin{figure}[t]
\centering
\resizebox{0.8\hsize}{!}{\includegraphics[width=7cm,height=4.5cm]{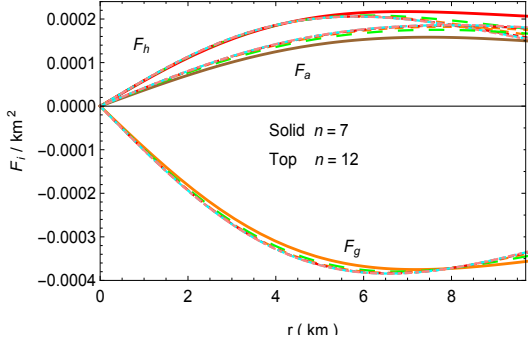}}
\caption{TOV-equation profile of PSR J1614-2230 for $M = 1.97 M_{\odot},R = 9.69 km$ and $b = 0.04$.}\label{fit}
\end{figure}

The above expression may also be written as
\begin{equation}
F_g+F_h+F_a=0,
\end{equation}
where $F_g, F_h$ and $F_a$ represents the gravitational, hydrostatics and anisotropic forces respectively and can be written as,
\begin{eqnarray}
F_g &=& -\frac{\nu'}{2}(\rho+p_r)\\
F_h &=& -\frac{dp_r}{dr}\\
F_a &=& {2\Delta \over r}.
\end{eqnarray}

The profile of three different forces are plotted in Fig. \ref{fit} and we can see that the system is in equilibrium state.

\subsection{Causality and stability condition}

In this section we are going to find the subliminal velocity of sound and stability condition. For a physically acceptable model of anisotropic fluid sphere the radial and transverse velocities of sound should be less than 1, which is known as the causality condition. The radial velocity $(v_{sr}^{2})$ and transverse velocity $(v_{st}^{2})$ of sound can be obtained as

\begin{eqnarray}
v_{r}^{2} = {dp_r \over d\rho}=\alpha~~,~~v_{t}^{2} = {dp_t \over d\rho}.
\end{eqnarray}

\begin{figure}[t]
\centering
\resizebox{0.8\hsize}{!}{\includegraphics[width=7cm,height=4.5cm]{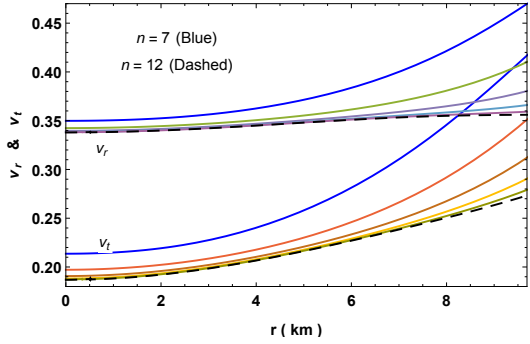}}
\caption{Velocity of sound profiles of PSR J1614-2230 for $M = 1.97 M_{\odot},R = 9.69 km$ and $b = 0.04$. }\label{fis}
\end{figure}

\begin{figure}[t]
\centering
\resizebox{0.8\hsize}{!}{\includegraphics[width=7cm,height=4.5cm]{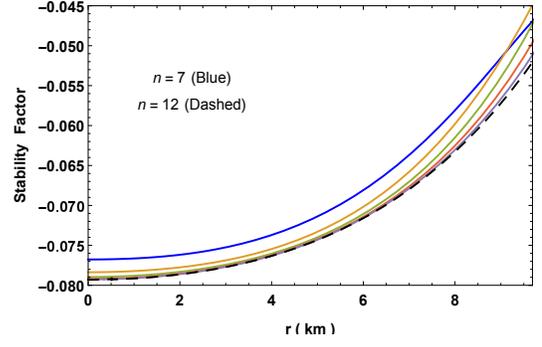}}
\caption{Stability factor ($v_t^2-v_r^2$) profiles of PSR J1614-2230 for $M = 1.97 M_{\odot},R = 9.69 km$ and $b = 0.04$.}\label{fist}
\end{figure}

The profile of radial and transverse velocities of sound have been plotted in Fig. \ref{fis}, the figure indicates that our model satisfies the causality condition. Now the stability condition proposed by Abreu \cite{abr07} i.e. $-1 \le v_t^2-v_r^2 \le 0$ (Fig. \ref{fist}).

\subsection{Adiabatic index and stability condition}

For a relativistic anisotropic sphere the stability is related to the adiabatic index $\Gamma$, the ratio of two specific heats, defined by \cite{cha93},
\begin{equation}
\Gamma_r=\frac{\rho+p_r}{p_r}\frac{dp_r}{d\rho}.
\end{equation}

\begin{figure}[t]
\centering
\resizebox{0.8\hsize}{!}{\includegraphics[width=7cm,height=4.5cm]{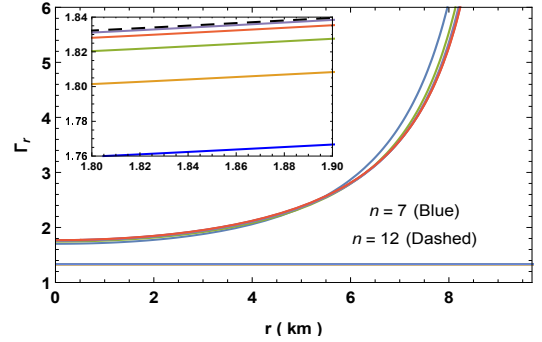}}
\caption{Adiabatic index profiles of PSR J1614-2230 for $M = 1.97 M_{\odot},R = 9.69 km$ and $b = 0.04$.}\label{fiad}
\end{figure}

Now $\Gamma_r>4/3$ gives the condition for the stability of a Newtonian sphere and $\Gamma =4/3$ being the condition for a neutral equilibrium proposed by \cite{bon64}. This condition changes for a relativistic isotropic sphere due to the regenerative effect of pressure, which renders the sphere more unstable. For an anisotropic general relativistic sphere the situation becomes more complicated, because the stability will depend on the type of anisotropy. For an anisotropic relativistic sphere the stability condition is given by \cite{cha93},

\begin{equation}
\Gamma>\frac{4}{3}+\left[\frac{4}{3}\frac{(p_{ti}-p_{ri})}{|p_{ri}^\prime|r}+\frac{8\pi}{3}\frac{\rho_ip_{ri}}{|p_{ri}^\prime|}r\right]_{max},
\end{equation}
where, $p_{ri}$, $p_{ti}$, and $\rho_i$ are the initial radial, tangential pressures and energy density in static equilibrium satisfying (\ref{to1}). The first and last term inside the square bracket represent the anisotropic and relativistic corrections respectively and both the quantities are positive that increase the unstable range of $\Gamma$ \cite{her92,cha93}. For this solution the adiabatic index is more than 4/3 and hence stable, Fig. \ref{fiad}.

\begin{figure}[t]
\centering
\resizebox{0.8\hsize}{!}{\includegraphics[width=7cm,height=4.5cm]{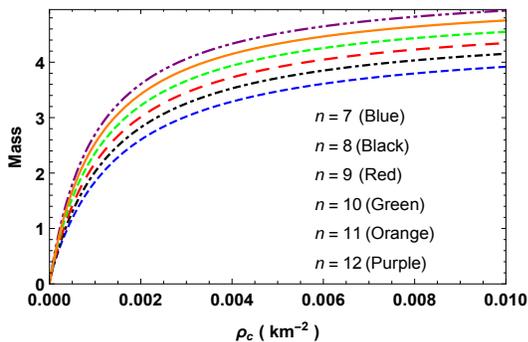}}
\caption{$M-\rho_c$ profiles with $R = 10.86 km$ and $b = 0.04$.}\label{fimr}
\end{figure}

\subsection{Harrison-Zeldovich-Novikov static stability criterion}

The stability analysis of Harrison et al. \cite{har65} and \cite{zel71} have shown that the adiabatic index of a pulsating star is same as in a slowly deformed matter. This leads to a stable configuration only if the mass of the star is increasing with central density i.e. $\partial m /\partial \rho_c > 0$ and unstable if $\partial m /\partial \rho_c < 0$.

In our solution, the mass as a function of central density can be written as
\begin{eqnarray}
m (\rho_c) &=& \frac{R}{2} \left(1-\frac{3b}{3b^n R^n+8\pi  b \rho_c  R^2+3b}\right) \label{mrhc}\\
{\partial m (\rho_c) \over \partial \rho_c} &=& \frac{12 \pi  b^2 R^3}{\left[3 b^n R^n+b \left(8 \pi  \rho  R^2+3\right)\right]^2}> 0.\\
\end{eqnarray}

The satisfaction of the above condition is shown as a plot in Fig. \ref{fimr}.

\begin{figure}[t]
\centering
\resizebox{0.8\hsize}{!}{\includegraphics[width=7cm,height=4.5cm]{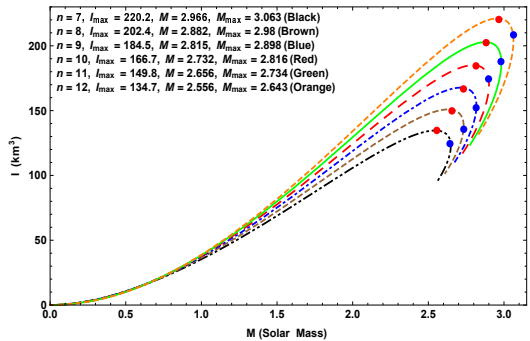}}
\caption{Variation of moment of inertia w.r.t. mass for $a=0.001$ and $b=0.04$. The red dots represents $\big(M,I_{max} \big)$ and blue dots $\big(M_{max},I \big)$}\label{im}
\end{figure}

\section{Discussion and conclusion}

The solution of Einstein's field equation with $e^{-\lambda} = 1+ar^2$ was presented by Duorah-Ray \cite{duo}, however, Finch-Skea \cite{} pointed out that the Duorah-Ray (DR) solution doesn't satisfy the field equations. Therefore, Finch-Skea (FS) corrected the solution and hence known as FS solution. FS not only corrected the DR solution but also performed extensive works to describe physically realistic neutron stars. The resulting equation of state from FS solution was also compared with Walecka's relativistic mean-field theory description and found to be quite in good agreement.  

An interesting result was presented by Bhar et al. \cite{pbh} showing that with the assumption of electric charge and Adler $g_{tt}$ metric potential in the Karmarkar condition, one leads to FS $g_{rr}$ metric potential which is a well behaved solution while its neutral counterpart isn't. 

The current paper generalized the FS $g_{rr}$ with the higher order term  $b^{n-1}r^n$. We also successfully analysed the behaviour of the solution showing its well behaved range w.r.t. the parameter $n$. It is found that the solution exist and satisfy causality condition for  $n=4,~5$ and within the range $7\le n \le 12$. All the solutions correspond to other values are not well-behaved. The fulfillment of the stable static criterion signifies that the solution is static and stable. The satisfaction of TOV-equation also implies the solution is in equilibrium. We have also plotted the M-R diagram for the range $7\le n \le 12$ and it shows that the maximum mass increases with $n$. For $n=7$ the maximum mass is 2.643$M_\odot$ with radius 8.976 km and for $n=12$, $M_{max} =3.063M_\odot$ with radius 10.85 km. The profile of adiabatic index (see Fig. 9) shows that the equation of state gets stiffer for larger values of $n$ since the central values of $\Gamma_r$ are larger. This increases the stiffness of the equation of state leading to increase the maximum mass.

The stiffness of an EoSf is also link with moment of inertia of the compact star. For a uniformly rotating star with angular velocity $\Omega$ the moment of inertia is given by \cite{latt}
\begin{eqnarray}
I = {8\pi \over 3} \int_0^R r^4 (\rho+p_r) e^{(\lambda-\nu)/2} ~{\omega \over \Omega}~dr
\end{eqnarray}

where, the rotational drag $\omega$ satisfy the Hartle's equation \cite{hart}
\begin{eqnarray}
{d \over dr} \left(r^4 j ~{d\omega \over dr} \right) =-4r^3\omega~ {dj \over dr} .
\end{eqnarray}
with $j=e^{-(\lambda+\nu)/2}$ which has boundary value $j(R)=1$. The approximate moment of inertia $I$ up to the maximum mass $M_{max}$ was given by Bejger and Haensel \cite{bejg} as
\begin{equation}
I = {2 \over 5} \Big(1+x\Big) {MR^2},
\end{equation}
where parameter $x = (M/R)\cdot km/M_\odot$. For the solution we have plotted mass vs $I$ in Fig. \ref{im} that shows as $n$ increases, the mass also increase and the moment of inertia increases till up to certain value of mass and then decreases. Therefore, we can say that as moment of inertia increases, the stiffness of the corresponding EoS also increases. Comparing Figs. \ref{fim} and \ref{im} we can see that the mass corresponding to $I_{max}$ is not equal to $M_{max}$ from $M-R$ diagram. In fact the mass corresponding to $I_{max}$ is lower by $\sim 3$\% from the $M_{max}$. This happens to the EoSs without any strong high-density softening due to hyperonization or phase transition to an exotic state \cite{bej}.


\end{document}